\documentclass[runningheads]{svmult}

\usepackage{makeidx}   
\usepackage{graphicx}  
\usepackage{subeqnar}  
\usepackage{multicol}  
\usepackage{physprbb}  
\makeindex             

\begin{document}
\title*{Stochastic Dynamics of Vortex Loop.
\protect\newline Large Scale Stirring Force }

\toctitle{Stochastic Dynamics of Vortex Loop. \protect\newline
Large Scale Stirring Force}

\titlerunning{Stochastic Dynamics of Vortex }
\author{S.K. Nemirovskii and A. Ja. Baltsevich}
\institute{Institute of Thermophysics,\ 630090, Novosibirsk,\
Lavrent'eva, 1.}

\maketitle

\begin{abstract}
Stochastic dynamics of a vortex filament obeying local induced
approximation equation plus random agitation is investigated by
analytical and numerical methods. The character of a stirring
force is supposed to be a white noise with spatial correlator
concentrated at large distances comparable with size of the loop.
Dependence of the spectral function $\left\langle \mathbf{s}
_\kappa ^\alpha \mathbf{s}_\kappa ^\beta \right\rangle $ of the
vortex line on both the one-dimensional wave vector $\kappa $ and
intensity of the external force correlator $\left\langle
\mathbf{\zeta }_\kappa ^\alpha \mathbf{\zeta }_\kappa ^\alpha
\right\rangle $ was studied. Here $\mathbf{s} _\kappa ^\alpha $ is
the Fourier transform of the line element position $
\mathbf{s}^\alpha (\xi ,t)$. It is shown that under the influence
of an external random force a vortex ring becomes a small tangle
whose mean size depends on external force intensity. The
theoretical predictions and the numerical results are in
reasonable agreement.
\end{abstract}

\section{Introduction}

In the previous paper\cite{previous} we discuss how the
large-scale perturbations can destroy the thermal equilibrium
state in the space of vortex loop configurations. In this paper we
elaborate that idea and present results of the both analytical and
numerical investigations on stochastic dynamics of a vortex
filament in HeII undergoing an action of the large scale random
displacements. Moreover we consider the case when the smooth
dissipation connected with normal component is small (that
correspond to the case of very small temperature ) and the only
strong dissipative mechanisms appear at very small scales
comparable with the core radius of vortex. Thus detailed (at which
scale) balance between the pumping and dissipation required for
themal equilibrium is violated and, as it has been discussed in
\cite{previous}, essentially nonequilibrium picture state must
develop.

From a formal point of view that problem is significantly more
involved, therefore we restrict ourselves to a consideration of
the local induction approximation \cite{Don_book},\cite{SCHWARZ}
in the equation of motion and omit processes of reconnection. This
statement of problem is, of course, far from real superfluid
turbulence in He~II. A value of that work is that it enables us to
understand mechanisms of entanglement of vortex filament and of
appearing the strongly nonequilibrium state. We would remind that
idea of the vortex tangle had been launched by Feynman more than
40 years ago \cite {Don_book} and only about 10-15 years ago
Schwarz demonstrated and confirmed that idea in his famous
numerical simulations\cite{SCHWARZ}. To our knowledge a similar
success in analytic study is absent.

A structure of the paper is following. In the first part of this
paper we develop nonequilibrium. diagram technique analogous to
the one elaborated by Wyld \cite{WYLD} for classical turbulence.
Using further method of direct interaction approximation we derive
a set of Dyson equation for the pair correlators and for the Green
functions. Assuming that region of stirring force and dissipation
are widely separated in space of scales we seek for a scale
invariant solution in the so called inertial interval. We also
present results of direct numerical simulation of the vortex
tangle dynamics. Numerical results confirm the ones obtained in
the analytical investigations, however some discrepancies
remained.

\section{Analytical Investigation}

In the local induction approximation (LIA) the equation of motion
of quantized vortex filament in HeII reads

\begin{equation}
{\frac{{d~}\mathbf{s}{(\xi
,t)}}{{d~t}}}~=~\tilde{\beta}\mathbf{s}~^{\prime }\times
\mathbf{s}~^{\prime \prime }~+\delta +\mathbf{\zeta }(\xi ,t).
\label{e:n1}
\end{equation}
Here $\mathbf{s}(\xi ,t)$ is a point of the filament labeled by
the Lagrangian variable $\xi ,~0~\leq ~\xi ~\leq ~2\pi ,$ which
coincides here with the arclength. The quantity $\delta $ stands
for dissipation, which is small for usual scales and large for
marginally small scales comparable with the core size $r_0$.
External Langevin force $\mathbf{\zeta }(\xi ,t)$ is supposed to
be Gaussian with correlator

\begin{equation}
\left\langle \zeta ^\alpha (\xi _1,t_1)\zeta ^\beta (\xi
_2,t_2)\right\rangle ~=~F^\alpha (\xi _1-\xi _2)~\delta
(t_1-t_2)~\delta ^{\alpha \beta },~\;\alpha ,\beta ~=~1,2,3~,
\label{e:n2}
\end{equation}
where $F^\alpha (\xi _1-\xi _2)$ is changing on the large scale of
order of the line length ($\sim 2\pi $). The quantity
$\tilde{\beta}$ is $\tilde{\beta
}~=~{\frac{\stackrel{\symbol{126}}{\kappa }}{{4\pi }}}\log {\frac
R{r_0}}$, with circulation $\stackrel{\symbol{126}}{\kappa }$ and
cutting parameters $ R $ (external size, e.g. averaged radius of
curvature) and $r_0$. For our numerical calculations we have
chosen $\stackrel{\symbol{126}}{\kappa } ~=~10^{-3}cm^2/s$. This
value corresponds to the case of superfluid helium.

In Fourier space equation (\ref{e:n2}) has the form

\begin{equation}
-i\omega s_q^\alpha ~=~\int \Gamma _{\kappa \kappa _1\kappa
_2}^{\alpha \beta \gamma }s_{q_1}^\beta s_{q_2}^\gamma \delta
(q-q_1-q_2)dq_1dq_2~+\delta _q+~\zeta _q^\alpha ~.  \label{e:n3}
\end{equation}
Here \textbf{$s$}$_q^\alpha $ is the spacial and temporal Fourier
component of $\mathbf{s}^\alpha (\xi ,t)$, defined as follows:

\begin{equation}
\mathbf{s}_q^\alpha ~=~\int \int \mathbf{s}^\alpha (\xi
,t)~e^{i(\omega t~-~\kappa \xi )}~dtd\xi ~.  \label{e:n4}
\end{equation}
The vertex $\Gamma _{\kappa \kappa _1\kappa _2}^{\alpha \beta
\gamma }$ responsible for nonlinear interaction has the form

\begin{equation}
\Gamma _{\kappa \kappa _1\kappa _2}^{\alpha \beta \gamma
}~=~{\frac{{i\tilde{ \beta}}}{{2\sqrt{2\pi }}}}\epsilon ^{\alpha
\beta \gamma }\kappa _1\kappa _2(\kappa _2-\kappa _1),~
\label{e:n5}
\end{equation}
where $\epsilon ^{\alpha \beta \gamma }$ it the antisymmetric unit
tensor. One can show the vertex $\Gamma _{\kappa \kappa _1\kappa
_2}^{\alpha \beta \gamma }$ to satisfy the so called Jacoby
identities

\begin{equation}
\left[ \kappa ^n~\Gamma _{\kappa \kappa _1\kappa _2}^{\alpha \beta
\gamma }~+~\kappa _2^n~\Gamma _{\kappa _2\kappa \kappa _1}^{\gamma
\alpha \beta }~+~\kappa _1^n~\Gamma _{\kappa _1\kappa _2\kappa
}^{\beta \gamma \alpha }\right] \delta (\kappa +\kappa _1+\kappa
_2)~=~0,~\;\;n=2,4.  \label{e:n6}
\end{equation}
This relations express are tightly connected with  the laws of
conservation of total length $L$ and curvature $K$
\begin{equation}
\,\,\,L~=~\int_0^{2\pi }~\mathbf{s}~^{\prime }\mathbf{s}~^{\prime
}~d\xi ~=const,\;\;\;\;\,K~=~\int_0^{2\pi }~\mathbf{s}~^{\prime
\prime }\mathbf{s} ~^{\prime \prime }~d\xi ~=const~.
\label{e:n6a}
\end{equation}
Conservation of these quantities is readely derived from either of
relations (\ref{e:n1}),(\ref{e:n3}). It is understood that
conservation law is held in absence of the both dissipation and
stirring force.

One of the regular approaches to describe random fields is based
on the Wyld diagram technique~\cite{WYLD}, originally developed to
study hydrodynamic turbulence. Following this technique we
introduce for the description of random processes the following
averages: the spectral density tensor (or correlator, or simply
spectrum) $S_q^{\alpha \beta }$ and the Green tensor $ G_q^{\alpha
\beta }$ (or simply Green function) which are defined by

\begin{equation}
S_q^{\alpha \beta }~\delta (q+q_1)~=~\left\langle
\mathbf{s}_q^\alpha \mathbf{s}_{q_1}^\beta \right\rangle ~,
\label{e:n9}
\end{equation}
\begin{equation}
G_q^{\alpha \beta }~\delta (q+q_1)~=~\left\langle {\frac{{\delta
s_q^\alpha } }{{\delta \zeta _{q_1}^\beta }}}\right\rangle ~.
\label{e:n10}
\end{equation}
Analysis of diagrams shows that due to the antisymmetry of tensor
$\epsilon ^{\alpha \beta \gamma }$ contained in the expression for
the vertex $\Gamma _{\kappa \kappa _1\kappa _2}^{\alpha \beta
\gamma }$ , both $S_q^{\alpha \beta }$ and $G_q^{\alpha \beta }$
are proportional to $\delta _{\alpha \beta }$, i.e. $S_q^{\alpha
\beta }~\equiv ~S_q^\alpha $ and $G_q^{\alpha \beta }~\equiv
~G_q^\alpha $ . Details of that technique are described in
\cite{npp}.

The renormalized quantities $S_q^\alpha $ and $G_q^\alpha $
(taking into account interactions) satisfy a Dyson set of diagram
equations:

\begin{equation}
G_q^\alpha ~=~^{\circ }G_q^\alpha ~+~^{\circ }G_q^\alpha \Sigma
_q^\alpha G_q^\alpha ~,  \label{e:n11}
\end{equation}
\begin{equation}
S_q^\alpha ~=~G_q^\alpha \left( F_q^\alpha ~+~\Phi _q^\alpha
\right) G_q^{\alpha \star }~.  \label{e:n12}
\end{equation}
Here $^{\circ }G_q^\alpha $ is the ``bare'' Green function which
is equal to $(\omega ~-~\delta _\kappa )^{-1}$. The mass operators
$\Phi _q^\alpha $ and $\Sigma _q^\alpha $ can be written in form
of diagram series: These series frequently used in nonequilibrium
processes have a standard form, explicit form of them is given in
\cite{WYLD},\cite{npp}.

\section{ Conservation Laws and Pair Correlators}

Dyson equations have shapes indicating a cumbersome handling,
therefore they can be studied for some special cases. One of them
is considered in the present paper. It is connected with
conservation laws for the total length and the curvature expressed
by (\ref{e:n6a}). Let us consider conservation of total curvature
(for total length there is the same consideration). In Fourier
space the conservation laws for total curvature $\kappa $ can be
expressed in the following form
\begin{equation}
{\frac{{\partial K_\kappa }}{{\partial t}}}~+~{\frac{{\partial
P_\kappa ^K}}{ {\partial \kappa }}}~=I_{+}^\kappa (\kappa
)~-~I_{-}^\kappa (\kappa ) \label{e:a1}
\end{equation}
where $K_\kappa ={\frac 1{\sqrt{2\pi }}}\int_0^{2\pi
}\mathbf{s}~^{\prime \prime }\mathbf{s}~^{\prime \prime
}e^{-i\kappa \xi }d\xi $ is the curvature density and $P_\kappa
^K$ is the flux of this quantity in Fourier space (or, equally, in
space of scales). The right-hand side of equation (\ref{e:a1})
describes creation of additional curvature (with rate
$I_{+}^K(\kappa )$) due to external force and annihilation of it
due to dissipative mechanism (with rate $-I_{-}^K(\kappa )$). In
the equilibrium case the flux $P_\kappa ^K$ is absent and source
and sink terms must compensate each other locally for each $\kappa
$, i.e. $I_{+}^K(\kappa )~=~I_{-}^\kappa (\kappa )$. In the case
under consideration when source and sink terms are widely
separated in $ \kappa $-space that condition is obviously
violated. Therefore a flux of curvature $P_\kappa ^K$ in Fourier
space appears. In region of wave numbers $ \kappa $ remote from
both region of the pumping $\kappa _{+}$ and of the sink $\kappa
_{-},\;\;\kappa _{+}~\ll ~\kappa ~\ll ~\kappa _{+}$ , the so
called inertial interval, derivative $\partial P_\kappa
^K/\partial \kappa =0 $ , so $P_\kappa ^K$ is constant equal ,say,
$P^\kappa $. Resuming we conclude that the problem reduces to
study the set of Dyson equation (\ref {e:n11})-(\ref{e:n12}) in
inertial interval under condition of constant flux of the
curvature. In this case $S_q^\alpha $ and $G_q^\alpha $ are
expected to be independent on the concrete type of both the source
and the sink but to be dependent on value of $P^K$. Furthermore,
the vertices $\Gamma _{\kappa \kappa _1\kappa _2}^{\alpha \beta
\gamma }$ are homogeneous functions of its arguments. This
property, as well as the condition $\kappa _{+}~\ll ~\kappa _{-}$
by virtue of which one can put $\kappa _{+}~=~0$ and $ \kappa
_{-}~=~\infty $, leads to the assumption that the problem is the
scale invariant, i.e.\ it has no characteristic scale for $\kappa
$. This suggests a power-law form of $S_q^\alpha $ and $G_q^\alpha
$

\begin{equation}
S_q^\alpha ~=~{\frac 1{{\kappa ^{r+p}}}}f\left[ {\frac \omega
{{\kappa ^r}}} \right] ~,~~~~~~G_q^\alpha ~=~{\frac 1{{\kappa
^r}}}g\left[ {\frac \omega {{ \kappa ^r}}}\right] ~.
\label{e:n15}
\end{equation}
Here both $f$ and $g$ are dimensionless functions of their
arguments. We aim now to find the scaling indices $r$ and $p$.

The first relation between indices $r$ and $p$ can be found from
an analysis of diagram series, claiming all terms to have the same
powers of argument $ \kappa $. This leads to the first scaling
condition

\begin{equation}
2r~+~p~=~7~.  \label{e:n16}
\end{equation}
Another relation between $r$ and $p$ can be obtained from the
Dyson equations (\ref{e:n11}), (\ref{e:n12}) which can be
rewritten in the form (see e.g.~\cite{LVOV})

\begin{equation}
\int d\omega ~Im\left\{ S_q^\alpha \Sigma _q^\alpha ~-~\Phi
_q^\alpha G_q^{\alpha \star }\right\} ~=~0~.  \label{e:n17}
\end{equation}
This relation plays the role of kinetic equations for systems with
a weak interaction . It has been obtained with help of the
expression for the Green function $G_q^\alpha ~=~(\omega ~-~\Sigma
_q^\alpha )^{-1}$; the external force correlator $F_\kappa ^\alpha
$ disappears in the inertial interval. To find a relation of
interest between $r$ and $p$ we rewrite relation ~(\ref {e:n17}),
disclosing expressions for mass operators $\Phi _q^\alpha $, $
\Sigma _q^\alpha $ and restricting ourselves to first order terms
in diagram series. That procedure called direct interaction
approximation is frequently used in classical turbulence (see
e.g.~\cite{LVOV}). After some calculation we arrive at the
following relation (see also \cite{KUZNETSOV}):

\begin{eqnarray}
Im\int ~d\omega d\omega _1d\omega _2d\kappa _1d\kappa _2\delta
(q+q_1+q_2)~\times \Gamma _{\kappa \kappa _1\kappa _2}^{\alpha
\beta \gamma } &&  \nonumber \\ \times \left\{ \Gamma _{\kappa
\kappa _1\kappa _2}^{\alpha \beta \gamma }G_q^\alpha S_{q_1}^\beta
S_{q_2}^\gamma ~+~\Gamma _{\kappa _2\kappa \kappa _1}^{\gamma
\alpha \beta }G_{q_2}^\gamma S_q^\alpha S_{q_1}^\beta ~+~\Gamma
_{\kappa _1\kappa _2\kappa }^{\beta \gamma \alpha }G_{q_1}^\beta
S_{q_2}^\gamma S_q^\alpha \right\} &~=~&0~.  \label{e:n21}
\end{eqnarray}
\begin{figure}[b]
\begin{center}
\includegraphics[width=.5\textwidth]{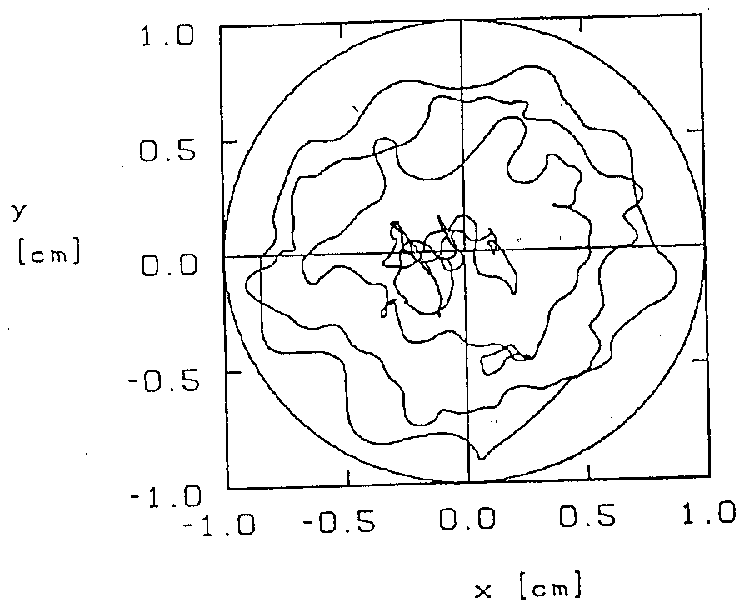}
\end{center}
\caption[]{} \label{f:1}
\end{figure}
\begin{figure}[b]
\begin{center}
\includegraphics[width=.5\textwidth]{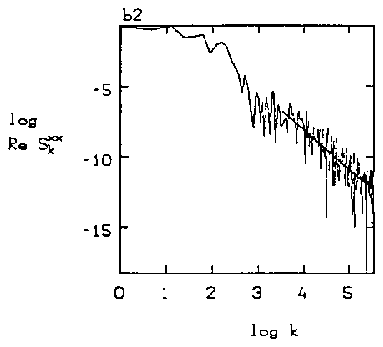}
\end{center}
\caption[]{} \label{f:2}
\end{figure}
To move further we perform conformal transformations in the second
and third term within the braces in integrand, known as Zakharov
transformations (see e.g. Zakharov~\cite{ZAKHAROV} and Kuznetsov
and L'vov ~\cite{KUZNETSOV}). For example for the second term
these transformations have the form

\begin{equation}
\kappa ~=~\kappa ~^{\prime \prime }(\kappa /\kappa ~^{\prime
\prime })~,~~\kappa _1~=~\kappa ~^{\prime }(\kappa /\kappa
~^{\prime \prime })~,~~\kappa _2~=~\kappa (\kappa /\kappa
~^{\prime \prime })~,  \label{e:n24}
\end{equation}
\begin{equation}
\omega ~=~\omega ~^{\prime \prime }(\kappa /\kappa ~^{\prime
\prime })^r~,~~\omega _1~=~\omega ~^{\prime }(\kappa /\kappa
~^{\prime \prime })^r~,~~\omega _2~=~\omega (\kappa /\kappa
~^{\prime \prime })^r~. \label{e:n25}
\end{equation}
The third term is transformed in similar manner. As a result the
integrand in (\ref{e:n21}) becomes

\begin{equation}
\Gamma _{\kappa \kappa _1\kappa _2}^{\alpha \beta \gamma
}G_q^\alpha S_{q_1}^\beta S_{q_2}^\gamma \left\{ \Gamma _{\kappa
\kappa _1\kappa _2}^{\alpha \beta \gamma }~+~\left[ {\frac \kappa
{\kappa _2}}\right] ^x\Gamma _{\kappa _2\kappa \kappa _1}^{\gamma
\alpha \beta }~+~\left[ {\frac \kappa {\kappa _1}}\right] ^x\Gamma
_{\kappa _1\kappa _2\kappa }^{\beta \gamma \alpha }\right\}
\label{e:n26}
\end{equation}
where

\begin{equation}
x~=~7~-~r~-~2p~.  \label{e:n27}
\end{equation}
Due to Jacoby identities (\ref{e:n6}) the integrand vanishes when
$x~=~-2$ for conservation of total length and $x~=~-4$~for
conservation of total curvature. Substiting  these values into
(\ref{e:n27}) and solving equations (\ref{e:n16}), (\ref{e:n27})
we obtaine a set of couples of indices $r$, $p$ corresponding to
nonequilibrium. states with fluxes of the length ($r~=~5/3$ ,
$p~=11/3$) and of the curvature ($r~=~1$, $p~=5$). One time
correlators can be found then integrating over frequences $\omega
$
\begin{equation}
S_\kappa ^\alpha ~=~\int ~d\omega {\frac 1{{\kappa
^{r+p}}}}f\left\{ {\frac \omega {{\kappa ^{\mathbf{s}}}}}\right\}
~\propto ~\left\{
\begin{array}{ll}
\kappa ^{-{\frac{11}3}} & \mbox{for length} \\ \kappa ^{-5} &
\mbox{for curvature}
\end{array}
\right.   \label{e:n28}
\end{equation}

So we have got solutions for the correlators $S_\kappa ^\alpha $
which correspond to different conservation laws in (\ref{e:n6}).
Since there are no sources and sinks acting in the intermediate
range these solutions guarantee that the according fluxes are
constant. Depending on the way of agitation of the system one can
get the real spectrum as some mixture of the obtained solutions in
which the fluxes of length and curvature are present
simultaneously. A similar situation for wave systems has been
discussed earlier (see e.g.\ \cite{LVOV} and bibliography therein)
and it is called multi-flux solution.

Having in mind to compare our result with the both numerical and
experimental investigation we have to take into account a presence
of $ \delta $-correlated (in $\xi $-space) random force considered
in our previous paper \cite{previous}. In the local induction
approximation the energy $H\{\mathbf{s\}}$ of line is proportional
of its length and in parametrization when $\xi $ is arclength. can
be expressed as
\begin{equation}
H\{\mathbf{s\}=}\frac{{\rho }_s{\stackrel{\symbol{126}}{\kappa
}}^2}{4\pi } \ln \frac R{r_0}\int\limits_\Gamma \mathbf{s}^{\prime
}(\xi )\mathbf{s} ^{\prime }(\xi )d\xi   \label{H_local}
\end{equation}
It is easy to see that the equilibrium distribution described in
\cite{previous} leads in that case to correlator $S_\kappa ^\alpha
$ $\propto 1/\kappa ^2.$
The final solution is a mix of
equilibrium solution and of the ones expressed by relation
(\ref{e:n28}). Because of nonlinearity it, in general, is not a
simple superposition except of the  cases when one of  stirring
action prevails and the other can be considered as small
deviations. For instance if a large-scale random stirring is small
in comparison with $ \delta $-correlated (in $\xi $-space) action
we have

\begin{equation}
S_\kappa ^\alpha ~=~{\frac{{A}}{{\kappa ^2}}+\frac{{B}}{{\kappa
^{11/3}}} +\frac C{{\kappa ^5}}.}  \label{S_final}
\end{equation}
The second and third terms in the right-hand side of \ref{S_final}
are small. The constants $A,B,C$ entering are connected with both
intensity of random stirring and its structure. The further
specification requires some additional analysis.

\section{Some Numerical Results}

In this section we present some preliminary results on a direct
numerical simulations of a vortex ring evolution under action of a
random stirring displacements. The large scale character of noise
was guaranteed by calculating it from a Fourier series taking into
account only the first few harmonics. Besides some (uncontrolled)
white noise due to numerical procedure has been excited.
Fig.~\ref{f:1} shows the projection of the line in the $x,y$ -
plane (where the ring was placed initially) for several times. As
predicted, an consequent arising of higher harmonics takes place
leading eventually to an entanglement of the initially smooth
vortex loop.

Another numerical results is shown in Fig.~\ref{f:2} where
logarithm of quantity $S_\kappa ^{xx}$ averaged over several
realizations is depicted as a functions of $\log \kappa .$ The
average slopes the graphs depend on intensity large-scale stirring
force. In several realizations the slope lies between $-2.5$ and
$-3.5$, which agrees with theoretical prediction \ref {S_final}.

This work was partly supported under grant N 99-02-16942 from
Russian Foundation of Basic research.

%


\begin{thebibliography}{9}
\bibitem{previous}  S.K. Nemirovskii, L.P. Kondaurova, M. Tsubota: {
Stochastic Dynamics of Vortex Loop. Thermal Equilibrium},
cond-mat/0011175.

\bibitem{Don_book}  R.J. Donnelly:\textit{Quantized Vortices in Helium II}
(Cambridge University Press, Cambridge 1991)

\bibitem{SCHWARZ}  K. W. Schwarz: Phys. Rev. \textbf{B 18, }245 (1978), K.
W. Schwarz, Phys. Rev. \textbf{B 38} 2398 (1988)

\bibitem{WYLD}  H. W. Wyld, Ann. Phys. \textbf{14, } 134 (1961)

\bibitem{npp} S.K. Nemirovskii, J. Pakleza, W. Poppe:Stochasti
behaviour of a vortex filament. Notes et Documents LIMSI N91-14,
Orsay (1991)


\bibitem{LVOV}  V. E. Zakharov, V. S. L'vov, G. Falkovich: \textit{Kolmogorov
Spectra of Turbulence I}, (Springer-Verlag, 1992)

\bibitem{ZAKHAROV}  V. E. Zakharov: Zh. Experim. Theor. Phys.\textbf{\ 51,}
688 (1966)

\bibitem{KUZNETSOV}  E. A. Kuznetsov, V. S. L'vov: Phys. Lett. \textbf{
64A,} 157 (1977)
\end{thebibliography}
\end{document}